\documentclass[aps,superscriptaddress,nofootinbib,twocolumn]{revtex4-1}
\usepackage{amsfonts} % para incluir fonts y
\usepackage{amssymb}  % simbolos de la AMS
\usepackage{graphicx} % standard LaTeX graphics tool
\usepackage{amsmath}  % for including eps-figure files
\usepackage{amsmath}  % paquete matemÃ¡tico
\begin{document}

\title{Quantized Faraday effect in (3+1)-dimensional and (2+1)-dimensional
    systems}

\author{L. Cruz Rodr\'iguez}
\email{lcruz@fisica.uh.cu}
\affiliation{Departamento de F\'{\i}sica General, Facultad de F\'{\i}sica, Universidad de la Habana, San L\'azaro y L, Vedado La Habana, 10400, Cuba}

\author{A. P\'erez Mart\'inez}
\email{aurora@icimaf.cu}
\affiliation{Instituto de Cibern\'{e}tica Matem\'{a}tica y F\'{\i}sica (ICIMAF) \\
Calle E esq 15 No. 309 Vedado, La Habana, 10400, Cuba}

\author{H. P\'erez Rojas}
\email{hugo@icimaf.cu}
\affiliation{Instituto de Cibern\'{e}tica Matem\'{a}tica y F\'{\i}sica (ICIMAF) \\
Calle E esq 15 No. 309 Vedado, La Habana, 10400, Cuba}

\author{ E. Rodr\'iguez Querts}
\email{elizabeth@icimaf.cu}
\affiliation{Instituto de Cibern\'{e}tica Matem\'{a}tica y F\'{\i}sica (ICIMAF) \\
Calle E esq 15 No. 309 Vedado, La Habana, 10400, Cuba}

\begin{abstract}
We study Faraday rotation in the quantum relativistic
limit. Starting from the photon self-energy  in the presence of a constant magnetic
field the rotation of the polarization vector of a plane electromagnetic wave which travel
along the  fermion-antifermion  gas is studied. The connection between Faraday Effect and Quantum Hall Effect
 (QHE) is discussed.
%The particular case of propagation along the magnetic field is considered.
The Faraday Effect is also investigated for a massless relativistic
(2D+1)-dimensional fermion system which is derived by using the
compactification along the dimension parallel to the magnetic field.
The Faraday angle shows a quantized behavior as Hall conductivity in
two and three dimensions.
\end{abstract}
\maketitle

\section{Introduction}
\label{sec1}

Plane-polarized light penetrating in a magnetized transparent
charged medium and moving parallel to the magnetic field \textbf{B}
rotates its plane of polarization  as a consequence of
birefringence: the incoming wave splits in two opposite circularly
polarized  modes moving with different speeds (and frequencies), and
the polarization vector rotates. This is the Faraday effect
\cite{Faraday}.

Faraday rotation (FR) is clearly manifest for photon propagation
parallel to the magnetic field. The symmetry  properties behind
theFaraday effect are the following: the field \textbf{B} (which we
take along the $x_3$ axis), breaks the Lorentz symmetry group in two
subgroups, the  translations along $x_3$ (leading to the
conservation of momentum component $p_3$), and the rotations around
$x_3$ (leading to the conservation of angular momentum $J_3$). The
generator of rotations around $J_3$ is the antisymmetric matrix
$A_{3ij}=-A_{3ji}=\delta_{i1}\delta_{j2}$ whose eigenvectors are
proportional to the complex unit vectors
$\mathbf{e}^{\pm}=(\mathbf{e}_1 \mp i \mathbf{e}_2)/\sqrt{2}$ ,
where $\mathbf{e}^{\pm}$ are related respectively to positive and
negative circular polarizations. If the system is $C$ invariant,
both opposite circular polarizations contribute symmetrically,
leading to equal speeds of light propagating along \textbf{B}. If
the system is noninvariant under $C$, the speeds of light differ for
opposite circular polarizations. The resulting polarization vector
rotates describing a circumference (in general an ellipse), and the
Faraday effect arises.

%\emph{In the previous discussion we have assumed \textbf{$k_3$} parallel to \textbf{B} thus positive and negative polarizations correspond to right (R) and left (L) photon helicities. If \textbf{$k_3$} is antiparallel to \textbf{B} the helicities are just the opposite.

For propagation perpendicular to \textbf{B}, two elliptically polarized modes arise, one of their semiaxes being along the propagation vector $k$, the rotation being in the plane orthogonal to \textbf{B} containing \textbf{k} \cite{1}-\cite{Elizabeth}.

The FR is indeed a particular case of the general problem of photon
propagation  in a charged medium \cite{hugo1982}. In nonrelativistic
media, like the ionosphere and insulators it is a well-known
phenomena where the classical and semiclassical approaches can be
applied successfully \cite{Faraday}.

However, FR effects have been also observed in electromagnetic waves
coming from astrophysical objects \cite{Olivo1-6}. Some of these
sources are compact objects which are characterized by high
densities and strong magnetic fields that can reach up to $10^{15}$
G in magnetars \cite{Duncan:2000pj}. Hence the physical process
involved in the case of compact objects requires an adequate
treatment from the point of view of a quantum-relativistic approach.

On the other hand, in quasiplanar condensed matter systems such as
graphene (a genuine monolayer of carbon atoms in a honeycomb array,
whose theoretical properties are essentially described by a
two-dimensional relativistic chiral fermion system
\cite{NovoselovNature2005}-\cite{aurora}),  FR is  observed when
light propagates perpendicular to the graphene layer in the presence
of a static magnetic field with $\textbf{k}\parallel \textbf{B}$ and
the relation between the Faraday angle and the nonstatic
($\omega\ne0$) Hall conductivity has been pointed out
\cite{japoneses}-\cite{castro}.

From the point of view of  novel applications of graphene,
magneto-optical  phenomena such as Faraday effect must be understood
from both the theoretical and experimental points of view.
Theoretical studies of the conductivity tensor in the static limit
\cite{aurora}, \cite{Gusyin}  and nonstatic regime have been done
\cite{gusynin2009}-\cite{gusyninnewjournal}. Recently, FR has been
detected in single-layer and multilayered epitaxial graphene
\cite{experimental}. The measurements report a giant value of the
rotation angle which comes exclusively from the graphene system (the
substrate did not show any FR).

In spite of the differences in contexts, the description of the
Faraday effect in both, the astrophysical and graphene scenarios,
can be theoretically tackled by considering photon propagation
parallel to the constant magnetic field in quantum-relativistic
dense matter.

The scope of the present paper is to describe the FR effect and to
obtain the Faraday  angle for (3+1)-dimensional (3D+1) and 2+1
dimensional (2D+1) systems, starting from  the same formalism: the
relativistic conductivity tensor in 3D+1 for a massive fermion
system. Our goal will be to show the connection between Hall
conductivity and the Faraday effect and the quantization of the
Faraday angle \cite{Volkov}.

In a previous paper \cite{aurora}, the Hall conductivity for a
massless relativistic fermion system was studied by starting from
the quantum-relativistic photon self-energy tensor in the QED
framework using the approach of Ref. \cite{aurora90} to obtain  the
static limit. Now, as we are interested in FR, this  problem should
be generalized to the nonstatic limit  ($\omega\neq 0$).

In Ref. \cite{hugo1982} a detailed study of the structure of the
photon self-energy in the presence of a magnetic field at finite
density and temperature was done.  Photon self-energy satisfies
properties of gauge and Lorentz and \textit{CPT} invariance. General
properties of the photon self-energy and the dispersion equations
for photons propagating in the medium were solved in  two cases:
photon propagating parallel  and perpendicular to the magnetic field
\cite{hugo1982}.

In this paper we  focus on the propagation parallel to the magnetic
field which establishes a relation between the Hall and Faraday
effects, so we take advantage of these calculations. We  have also
particularized the study to  2D+1 with the aim to describe
graphene-like systems.

Our calculations have been done in the imaginary-time formalism. For
$3D+1$ and $2D+1$ systems we have obtained the FR angle that
 the light undergoes upon propagating in a dense fermion system where the
chemical potential $\mu$ is greater than temperature $T$ ($\mu\gg
T$).

The weak-field limit for light propagating in a magnetized plasma
was  studied in \cite{Avjit}, taking  the dependence of the
self-energy with regard to \textbf{B} in a linear approximation. In
Ref. \cite{olivo} a calculation of the photon self-energy is made
for strong and moderate fields but $\mu \gg eB$. In both papers the
Faraday effect is considered in some particular cases and  the
semiclassical results have been reproduced. The real-time formalism
was used in Refs. \cite{Avjit}-\cite{olivo}.

Our findings are  relevant for two main reasons: first, we have
extended them to the  nonstatic limit from previous calculation
\cite{aurora90}-\cite{aurora} of 3D+1 relativistic massive fermions
and 2D+1 relativistic massless Hall and Ohm conductivities. Second,
we have found the connection between FR angle and Hall conductivity
(the Faraday angle depends on the admittivity: complex conductivity,
but the leading term is proportional to the Hall conductivity). This
result for 3D+1 and 2D+1 systems shows that it is a consequence of
general properties of QED in external magnetic fields at finite
density. The angle as a function of the photon frequency $\omega$
has branching points for 3D+1 relativistic dense massive fermion
systems as well as for 2D+1 massless systems a discrete set of
$\omega$ values. Hence, FR shows the effect of quantization of the
quantum Hall effect at nonzero frequency.  Our results for 2D+1
massless system are in agreement  with  previous theoretical work
for FR in graphene reported in Refs. \cite{Tan}-\cite{castro}. The
2D+1 results have been obtained from 3D+1 results by dimensional
compactification.

Astrophysical applications of our findings on FR angle  can be
expected in the context of radiation propagating through neutron
stars magnetospheres. This problem   will be discussed  in a
forthcoming presentation.

The paper is organized  as follows: In Sec. \ref{sec2} we start from
the one-loop approximation of the photon self-energy  in the
presence of a constant and uniform magnetic field and obtain the
relativistic Hall and Ohm conductivities in the nonstatic
approximation by generalizing the results obtained in Ref.
\cite{aurora90}. In Sec. \ref{sec4} the 3D+1 Faraday effect is
discussed and the Faraday angle is obtained to first order as half
of the Hall conductivity. Then, the 2D+1 massless QED limit is
obtained, and the expression for the Hall and Ohm conductivities are
written in Sec. \ref{secrel}. In Sec. \ref{sec5} the Faraday effect
and angle are discussed in 2D+1 dimensions obtaining the same
dependence with regard to the Hall conductivity as the 3D+1 case in
the first-order approximation. Finally, in Sec. \ref{sec6} we state
the concluding remarks. Appendices show the calculations relevant to
our results.

 \section{Photon self-energy in presence of magnetic field}
\label{sec2}
 The photon self-energy  in quantum electrodynamics in an
    external magnetic field was calculated at finite temperature and nonzero density in Ref. \cite{1}.
     The total electromagnetic field is written $A_{\mu}^e+a_{\mu}$ where $A_{\mu}^e$ refers to the
     external magnetic field and $a_{\mu}$ to the radiation field.  The photon self-energy (also called the
      polarization operator), can also be  interpreted as the linear-term coefficient of the functional expansion of the four-current $j_{\mu}$ in powers of the electromagnetic field $a_{\mu}$. That is, $j_{\mu}=\Pi_{\mu \nu}a_{\nu}$.

 The introduction of a chemical potential $\mu \neq 0$ is associated with a non-neutral electron-positron charged medium.
 The system is thus assumed as $C$ noninvariant, and total charge neutrality is guaranteed by the
 assumption of a hadron background. This background,  however,  is not taken into account in any of the further calculations.

 The generalized Furry's theorem \cite{Fradkin} establishes that odd
 powers of $\mu$ will be associated with odd powers of $A_{\mu}^e$ through
 antisymmetric tensor structures. That is, the self-energy contains antisymmetric odd-in-$\mu$ terms. Also,
  as gauge invariance is satisfied,  it implies that the self-energy tensor satisfies the four-dimensional gauge invariance condition
   $\Pi_{\mu\nu}k_{\nu}=k_{\mu}\Pi_{\mu\nu} =0$.
  We have for the photon self-energy the expression\footnote{Unless specified otherwise, we use natural units $\hbar=c$.}
  \begin{equation}
   \Pi_{\mu\nu}(x,y)=e^{2}Tr\hspace{-0.1cm}\int\hspace{-0.15cm}\gamma_{\mu}G(x,z)\Gamma_{\nu}(z,y',y)G(y',x)d^{4}zd^{4}y'.  \label{operador}
 \end{equation}
For the calculation of the components of $\Pi_{\mu\nu}$ we take (\ref{operador}) in the
 one loop approximation with the temperature Green's functions $G(x,y|A^e)$ being the solution
 of the Dirac equation in a constant magnetic field \textbf{B} such that $A_{\nu}^e=B x_{1}\delta_{\nu,2}$,
 directed along the $x_3$-axis
\begin{equation}
  [\gamma_{\nu}(\partial_{\nu}-ieA_{\nu}^e)+m]G(x,y|A^e)=\delta(x-y),  \label{diracequation}
 \end{equation}
\noindent where $\partial_{4}=\partial/\partial x_{4}-\mu$, and $\mu$ is the
  chemical potential for the electron-positron gas. It is also
  important to remark that Eq. (\ref{diracequation}) defines the fermion temperature-dependent Green's
  function for $x_{4}$ in the interval $-\beta$ to $\beta$; $\beta=1/T$.
  In the one-loop approximation the Fourier transform of the
   polarization tensor has the form
 %\begin{widetext}
   \begin{equation}
 \hspace{-0.22cm}  \Pi_{\nu\rho}(k_{4},\hspace{-0.06cm}\vec{x},\hspace{-0.06cm}\vec{x'}\hspace{-0.06cm}|\hspace{-0.08cm}A^{\hspace{-0.06cm}e}\hspace{-0.06cm})\hspace{-0.08cm}=\hspace{-0.08cm}\frac{e^{2}\hspace{-0.17cm}}{\hspace{-0.17cm}\beta}Tr\hspace{-0.15cm}\sum_{p_4}\hspace{-0.15cm}\gamma_{\nu}G(p_{4},\vec{x},\vec{x'}|\hspace{-0.08cm}A^e\hspace{-0.1cm})
   \gamma_{\rho}\hspace{-0.04cm}G(p_{4}\hspace{-0.03cm}+\hspace{-0.03cm}k_{4},\hspace{-0.07cm}\vec{x},\hspace{-0.07cm}\vec{x'}\hspace{-0.08cm}|\hspace{-0.07cm}A^{\hspace{-0.06cm}e}\hspace{-0.07cm}),
   \end{equation}
%\end{widetext}
   \noindent where $p_{4}=(2s+1)\pi/\beta$ and $s$ runs over integers from $-\infty$ to
   $+\infty$.
   Substituting the expression of the Green function of fermions $G(p_{4},\vec{x},\vec{x'}|A^e)$ we obtain
    the self-energy tensor as
 \begin{equation}
 \hspace{-0.22cm}
        \Pi_{\nu\rho}(k|A^{\hspace{-0.06cm}e}\hspace{-0.08cm},\mu,\beta^{-1})\hspace{-0.08cm}=\hspace{-0.08cm}\frac{e^{3}B}{2\pi^2\beta}\hspace{-0.1cm}\sum_{p_{4}}\hspace{-0.1cm}\sum_
        {n,n^{\prime}}\hspace{-0.1cm}\int\hspace{-0.15cm}\frac{dp_{3} C_{\nu,\rho}}{[p_{4}^{*}\hspace{-0.08cm}+\hspace{-0.08cm}\epsilon_{p,n}^{2}][(p_{4}^{*}+k_{4})\hspace{-0.08cm}+\hspace{-0.08cm}\epsilon_{p^{*}}^{2}]}
       ,\label{polarization1}
    \end{equation}
   \noindent where $p_{4}^{*}=p_{4}+i\mu$,  $\varepsilon_{p,n}=\sqrt{p_{3}^2+m^2+2enB}$ $\varepsilon_{p,n^{\prime}}=\sqrt{(p_{3}+k_{3})^2+m^2+2en^{\prime}B}$, n and $n^{\prime}$ are the Landau numbers and in what follows we will use the notation  $k_{\perp}^2\equiv k_{1}^{2}+k_{2}^{2}$  and $k_{\parallel}^2=k_{3}^{2}+k_{4}^{2}.$
Let us remark that the presence of the magnetic field in the $x_3$
direction breaks the spatial symmetry, hence
 in  Eq. (\ref{polarization1}) only the integral over $dp_{3}$ has survived. The integral over  $\int dp_{\perp}\rightarrow \sum_{n}\frac{\alpha_n eB}{(2\pi)^2}$ where $\alpha_n=2-\delta_{n0}.$

As  mention earlier, in Ref. \cite{hugo1982} the structure of
self-energy of the photon (\ref{polarization1}) in presence of a
magnetic field at finite density and temperature was obtained by
considering the  properties of  gauge and Lorentz invariance and
$CPT$ invariance. Six independent transverse tensors can be built in
terms of the four vectors: the momentum of the photon $k_{\mu}$, the
product of the external electromagnetic field tensor $F_{\mu\rho}$
and its square by $k_{\mu}$, leading respectively as
$F_{\mu\rho}k_{\rho}$ and $F_{\mu,\rho}^2k_{\rho}$, and the
four-velocity of the medium $u_{\mu}$ (a summary of these properties
can be found in the appendix). As our goal is to study the Faraday
effect in connection to the conductivity tensor  we will concentrate
on the case of a photon propagating parallel to the magnetic field
$(k_\perp=0)$ \cite{hugo1982}; in that case only three scalars are
independent. As we will show in the next sections only two of them
are related to the conductivity and also to Faraday effect.

 \subsection*{\textbf{Relativistic Hall and Ohm conductivity in non-static limit ($\omega\neq 0$)}}

This section is devoted to studying the 3D+1 relativistic Hall and
Ohm conductivities in the nonstatic limit ($\omega\neq 0$). The
expression for the spatial part of the current density is linear in
terms of the perturbative magnetic field and is given in terms of
the photon self-energy of the medium by \cite{aurora90}
        \begin{equation}
            j_{i}=\Pi_{i\nu}a_{\nu}, \ \ i=1,2,3, \,  \, \nu=1,2,3,4
        \label{eq:perturbativeconductivity}\end{equation}
        \noindent where $a_{4}=ia_{0}$ and $k_{4}=i\omega$, having in mind the transversality
        condition given by $\Pi_{\mu\nu}(k)k_{\nu}=0$, due to gauge invariance, Eq. (\ref{eq:perturbativeconductivity}) can be written as
\begin{equation}
            j_{i}=Y_{ij}E_{j}, \ \ i=1,2,3, \,\, j=1,2,3
        \label{eq:spatialcurrentdensity}\end{equation}
\noindent where $Y_{ij}=\Pi_{ij}/i\omega$ is the admittivity (complex conductivity tensor) and $E_{j}=i(\omega a_{j}-k_{j}a_{0})$ is
the electric field.

We will be especially interested in the real conductivity
$\sigma_{ij}=\text{Re}[Y_{ij}]$. The contribution to the current
density in Eq. (\ref{eq:spatialcurrentdensity}) due to $\sigma_{ij}$
can be written as
\begin{equation}
  j_{i}=\sigma_{ij}^{0}E_{j}+(E\times S)_{i},
\end{equation}
\noindent where
$\sigma_{ij}^{0}=\frac{\text{Im}[\Pi_{ij}^{S}]}{\omega}$ and
$S_{i}=\frac{1}{2}\epsilon^{ijk}\sigma_{jk}^{H}$ is a pseudovector
associated with
$\sigma_{ij}^{H}=\frac{\text{Im}[\Pi_{ij}^{A}]}{\omega}$. \noindent
$\Pi_{ij}^{A}$ and $\Pi_{ij}^{S}$ are, respectively, the
antisymmetric and symmetric parts of the polarization tensor
\cite{aurora} and \cite{aurora90}.

In the particular case where the electric field is the polarization
 vector of a transverse wave propagating along the magnetic field $\textbf{B}$,
 being $\textbf{E} \perp\textbf{B}$,  the conductivity tensor can be written in the following
 way:
\begin{equation}
\sigma_{ij}=\sigma^{0}\delta_{ij}+\epsilon_{ij}\sigma^{H},
\label{eq:conductividad}\end{equation}

\noindent where $\epsilon_{ij}$ is the antisymmetric $2\times 2$ unity tensor,
$\epsilon_{12}=-\epsilon_{21}=1$ and
\begin{equation}
\sigma^{0}=Im[t]/\omega,\label{sohm}
\end{equation}
\begin{equation}
\sigma^{H}=Im[r]/\omega,\label{shall}
\end{equation}
\noindent where the scalar quantities $r$ and $t$ depend on the
frequency- $\omega$, the momentum $k_3$ and also the temperature,
chemical potential and magnetic field \cite{hugo1982}. From  Eq.
(\ref{eq:conductividad}) we can identify  $\sigma^{0}$, $\sigma^{H}$
with the Ohm and Hall conductivities respectively. The scalar $r$
can be written as
\begin{equation}
r(k_{\parallel},\mu,B,T)=iI_r,\label{integralIr}
\end{equation}
\noindent and  $I_r$ is the integral
\[
I_r=\frac{e^3B\omega}{2\pi^2}\sqrt{\frac{k^2}{k_{\parallel}^2}}\sum_{n,n^{\prime}}F_{n\,n^{\prime}}^{(3)}(0)
\int\limits_{-\infty}^{\infty}\frac{dp_3(k_\parallel^2+2eB(n+n^{\prime}))}{D}
\]
\begin{equation}
\times(n_e(\varepsilon_{p,n})-n_p(\varepsilon_{p,n})),\label{intir}
\end{equation}

\begin{equation}
%\left|D\right|^{2}=
D=[2p_{3}k_{3}+k_{\parallel}^2+2eB(n^\prime-n)]^{2}-4\omega^{2}\varepsilon_{p,n}^{2}.\quad %\phi_{n,n^\prime}=k_{\parallel}^2+2eB(n+n^\prime)
\end{equation}
\noindent for $k_{\perp}\sim0$,  $\sqrt{\frac{k^2}{k_{\parallel}}}\rightarrow 1$,
$n_{e,p}(\varepsilon_{p,n})=(1+e^{(\varepsilon_{p,n}\mp \mu)\beta})^{-1}$ are the Fermi-Dirac distribution
 for fermions and antifermions, respectively.
\noindent The scalar $t$ is
\begin{equation}
t(k_{\parallel},\mu,B,T)=-\frac{e^3B}{4\pi^2}\sum_{n,n^{\prime}}F_{n\,n^{\prime}}^{(2)}(0)I_t,\label{tgeneral}
\end{equation}
\noindent where $F_{n\,n^{\prime}}^{(2,3)}(0)=\delta_{n,n^{\prime}-1} \pm \delta_{n-1,n^{\prime}}$  and $I_t$
\[
I_t=\int\limits_{-\infty}^{\infty}\frac{dp_3}{\varepsilon_{n,p}} (1-\frac{(2p_3k_3+J _{nn^{\prime}})(k_\parallel^{2}+2eB(n+n^\prime))}{D})
\]
\begin{equation}
\times (n_e(\varepsilon_{p,n})+n_p(\varepsilon_{p,n}))\label{integralIt}
\end{equation}
\noindent and $J_{nn^{\prime}}=k_{\parallel}^2+2eB(n^{\prime}-n).$

Thus, we have the expressions for the scalars r and t in the one
loop approximation for the fermion-antifermion gas, with the
assumption of $k_{\perp}=0$.  Now, starting from them we can study
the Hall and Ohm conductivities in some particular limits which are
relevant for applications in astrophysics as well as in
graphene-like systems.
\subsection{3D+1 Hall conductivity $\omega\neq0$}

The Hall and  Ohm conductivities Eqs. (\ref{sohm}) and (\ref{shall})
are given as imaginary parts of the scalars $r$ and $t$
respectively. At frequencies different from zero the integrals
(\ref{intir}) and (\ref{integralIt}) have singularities which come
from the zeros of the denominator $D$. But the Hall conductivity is
the imaginary part of $r=iI_r$, thus  the contribution to the Hall
conductivity comes from the real part of $I_r$, which means to
consider the  principal value of the integral [see the appendix,
first term of (\ref{delta})].  The result is the following:
\begin{equation}
Im[r(k_{\parallel},B,\mu,T)]=Re[I_r],
\end{equation}
\noindent where
\[
Re[I_r]=P(\frac{e^{3}B\omega}{2\pi^{2}}\underset{n,n^{\prime}}{\sum}F_{n,n}^{(3)}(0)\int\limits_{-\infty}^{\infty}dp_{3}\frac{
(k_{\parallel}^2+2eB(n+n^{\prime}))}{D}
\]
\begin{equation}
\times(n_{e}(\varepsilon_{p,n})-n_{p}(\varepsilon_{p,n}))).\label{eq:Imr}
\end{equation}
The denominator in (\ref{eq:Imr}) can be written as
$D\hspace{-0.1cm}=\hspace{-0.1cm}4k_\parallel^2(p_3-p_3^{(1)})(p_3-p_3^{(2)})$, where $p_3^{(1,2)}$ are the
 roots of the equation $D=0$ (see details in \cite{hugo1982})
 \begin{equation}
   p_3^{(1,2)}=\frac{-k_3J_{nn^{\prime}}\pm\omega\Lambda}{2k_\parallel^2},
 \end{equation}
and
\noindent
\begin{equation}
\Lambda=(J_{nn^{\prime}}^{2}+4\varepsilon_{0,n}^{2}k_\parallel^2)^{1/2}, \label{detalles1}
\end{equation}
and we have for $\sigma_{H}^{3D}$
 \begin{equation}\label{3Dhall}
 \hspace{-0.2cm}
\sigma_{H}^{3D}(k_\parallel,B,\mu,T)=\frac{\text{Im}[r]}{\omega}=\frac{Re[I_r(k_\parallel,B,\mu,T)]}{\omega},
\end{equation}
In the degenerate limit where $\mu\gg T$ and  $n_{e}(\epsilon_{p,n})$ are replaced by
a step functions $\theta(\mu-\varepsilon_{p,n})$, $n_{p}(\varepsilon_{p,n})\rightarrow 0$, after integration we obtain
\begin{widetext}
  \begin{equation}
  \sigma_{H}^{3D}(k_\parallel,B,\mu,T)=-\frac{e^{3}B}{2\pi^{2}}\sum_{n,n^{\prime}}^{n_{\mu},n_{\mu}^{\prime}}
  F_{n,n^{\prime}}^{(3)}(0)\frac{k_{\parallel}^2+2eB(n+n^\prime)}{4 \omega\Lambda}(\ln{|\frac{p_f-p_3^{(2)}}{p_f+p_3^{(2)}}|}+\ln{|\frac{p_f+p_3^{(1)}}{p_f-p_3^{(1)}}|}),
  \end{equation}
\end{widetext}
\noindent where $p_f=\sqrt{\mu^2-m^2-2neB}$ is the Fermi momentum.
As $F_{n,n^{\prime}}^{(3)}(0)$ are given by Kronecker $\delta$
expressions and taking the long wave limit $k_3\rightarrow 0$ we
have for $\sigma_{H}^{3D}$  the expression
\begin{widetext}
\begin{equation}
\sigma_{H}^{3D}(\omega,B,\mu,0)=\frac{e^{3}B}{2 \pi^2}(\underset{n=0}{\overset{n_\mu}{\sum}}\frac{\omega^2-2eB(2n+1)}{4m_n \omega^2}\frac{1}{\sqrt{(\frac{2eB-\omega^2}{2\omega m_n})^2-1}}\ln{\mid\frac{p_f/m_n-\sqrt{(\frac{2eB-\omega^2}{2\omega m_n})^2-1}}{p_f/m_n+\sqrt{(\frac{2eB-\omega^2}{2\omega m_n})^2-1}}\mid}\label{eq:3Dhallintegradaparan1}
\end{equation}
\begin{equation}
-\underset{n=1}{\overset{n_\mu}{\sum}}\frac{\omega^2-2eB(2n-1)}{4m_n \omega^2}\frac{1}{\sqrt{(\frac{2eB+\omega^2}{2\omega m_n})^2-1}}\ln{\mid\frac{p_f/m_n-\sqrt{(\frac{2eB+\omega^2}{2\omega m_n})^2-1}}{p_f/m_n+\sqrt{(\frac{2eB+\omega^2}{2\omega m_n})^2-1}}\mid}).\nonumber
\end{equation}
\end{widetext}
\noindent where $m_n\hspace{-0.1cm}=\hspace{-0.1cm}\sqrt{m^2+2neB}$.
From Eq. (\ref{eq:Imr}) at zero frequency  $\omega=0$ (static limit)
we recover the quantum Hall conductivity obtained in Ref.
\cite{aurora90}:\footnote{we have returned to the units $\hbar$ and
$c$ for obtain this result.}
\begin{equation}\label{sigmaHestatico}
 \sigma_{H}^{3D}(0,B,\mu,T)\hspace{-0.07cm}=\hspace{-0.07cm}\frac{e^{2}\hspace{-0.07cm}}{\hspace{-0.07cm}h^2}\hspace{-0.07cm}\sum_n^ {n_{\mu}}\hspace{-0.07cm}\alpha_n\hspace{-0.2cm}\int\limits_{-\infty}^{\infty}\hspace{-0.2cm}dp_{3}\theta(\mu-\varepsilon_{p,n})\hspace{-0.07cm}=\hspace{-0.07cm}\frac{e^{2}}{ch^2}\sum_n^ {n_{\mu}}\hspace{-0.1cm}\alpha_n p_f.
  \end{equation}
The sum over Landau levels is up to the integer number
$n_{\mu}=I[(\mu^2-m^2)/(2eB)]$, $n_{\mu}^{\prime}=n_{\mu}+1$.

\begin{figure}[h]
\begin{center}
 \includegraphics[width=8cm]{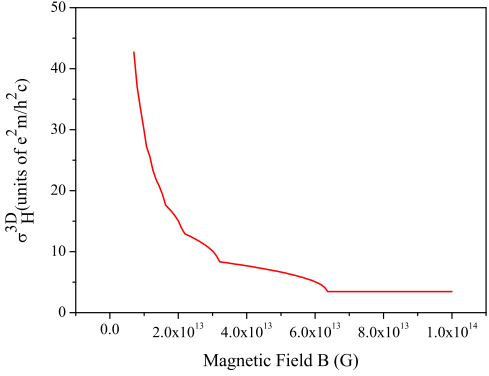}
 \caption{(Color online)3D Hall conductivity as a function of the magnetic field,
 where B runs between $7\times 10^{12}$ and $10^{14}$ G and $\hbar\omega=10^{-2} $ MeV.}\label{graficohall3dB}
\end{center}
\end{figure}

In Fig. \ref{graficohall3dB}, the three-dimensional (3D) nonstatic
Hall conductivity is plotted for constant chemical potential and
frequency as a function of the magnetic field. The curved step
behavior is illustrated; this behavior is also observed in the
static limit \cite{aurora90}.

\subsection{3D+1 Ohm conductivity $\omega\neq 0$}

Our aim now is to calculate the Ohm conductivity given by Eq.
(\ref{sohm}). A detailed calculation of  $Im[t]$ can be found in
\cite{hugo1982}. It expression comes from the imaginary part of the
integral  (\ref{integralIt}), which is related to the singularities
due to absorptive process, and it can be written in two different
cases, the first one, where $k_{\parallel}^2>0$, and absorption is
only due to excitation of particles, and a second one, where
$k_{\parallel}^2<0$ and absorption is due to excitation and also
pair creation. In the present study  we are going to use the
expression of $Im[t]$  in the region of $k_{\parallel}^2<0$ because
we will take the long wavelength limit ($k_3\longrightarrow 0$).
Furthermore, only the region of real frequencies, which means,
$k_3^{2}>k_\parallel^{2}$ is considered. To find the imaginary part
of $I_t$ the formulas (\ref{denominador})-(\ref{resultado}) will be
used, after that we have the Ohm conductivity as
 $$\sigma^{3D}_0(\omega,B,\mu,T)\hspace{-0.1cm}=\hspace{-0.1cm}\frac{Im[t]}{\omega}$$
 \[\hspace{-0.1cm}=\hspace{-0.1cm}\frac{e^{3}B}{8\pi\omega}\hspace{-0.1cm}\sum_{nn^{\prime}}\hspace{-0.1cm}\frac{\hspace{-0.1cm}F_{n,n^{\prime}}^{(2)}(0)
(k_{\parallel}^{2}+2eB(n+n^{\prime}))\hspace{-0.1cm}}
{\hspace{-0.07cm}{((k_{\parallel}^{2}+2eB(n^{\prime}\hspace{-0.07cm}-\hspace{-0.07cm}n))^2+4\varepsilon_{0,n}^2k_{\parallel}^{2})^{1/2}}\hspace{-0.07cm}}
\]
\[
\times\left(\theta(k_{\parallel}^{2}\hspace{-0.07cm}-\hspace{-0.07cm}k_{\parallel}^{2\,{\prime\prime}})\right.
\left[N(\varepsilon_{p}^{(m)})\hspace{-0.07cm}-\hspace{-0.07cm}N(\varepsilon_{p}^{(m)}\hspace{-0.07cm}+\hspace{-0.07cm}\omega)\right]
\]
\begin{equation}
+\left.\theta(k_{\parallel}^{2\,\prime}\hspace{-0.07cm}-\hspace{-0.07cm}k_{\parallel}^{2})\left[H(\varepsilon_{p}^{(j)})
+H(\omega\hspace{-0.07cm}-\hspace{-0.07cm}\varepsilon_{p}^{(j)})-2\right]\right)
\theta(k_{3}^{2}\hspace{-0.07cm}-\hspace{-0.07cm}k_{\parallel}^{2}),
\label{eq:Imtcompleto}\end{equation}
\noindent where
\[
 N(\varepsilon_{p,n})=n_e(\varepsilon_{p,n})+ n_p(\varepsilon_{p,n})\]
 \begin{align}
 H(\varepsilon_{p,n})=n_e(\varepsilon_{p,n})+ n_p(\omega-\varepsilon_{p,n}).\label{otros}
 \end{align}

\noindent Equation (\ref{eq:Imtcompleto}) is written for $\omega>0$,
where $\varepsilon_{p}^{(m)}$ and $\varepsilon_{p}^{(j)}$ are the
values of the  energy at the branching points for the excitation and
pair creation absorption processes respectively,
\[
\varepsilon_{p}^{(m)}=\frac{-\omega J_{nn^{\prime}}\pm k_{3}\Lambda}{2k_{\parallel}^{2}}, \quad m=(1,2),\]
\begin{equation}
\varepsilon_{p}^{(j)}=\frac{\omega J_{nn^{\prime}}\mp k_{3}\Lambda}{2k_{\parallel}^{2}}, \quad j=(3,4)\label{energia}
\end{equation}

The step function over $k_{\parallel}^2$ in Eq.
(\ref{eq:Imtcompleto}) defines the regions where excitation and pair
creation take place, where $k_{\parallel}^{2\,\prime}$ and
$k_{\parallel}^{2\,\prime\prime}$ are the branching points located
at $k_{\parallel}^{2}$ plane
 \noindent

\begin{equation}
k_{\parallel}^{2\,\prime}=-(\varepsilon_{0,n}+\varepsilon_{0,n^{\prime}})^{2} , k_{\parallel}^{2\,\prime\prime}=-(\varepsilon_{0,n}-\varepsilon_{0,n^{\prime}})^{2}.\label{detalles2}
\end{equation}

Now, our attention is focused in the long wavelength and degenerate limit, in order to get a better
understanding of our results, we separate them for each region.\\
 \textbf{a)}\textbf{ Region I}: $k_{\parallel}^{2}>k_{\parallel}^{2\,{\prime\prime}}$ excitation case.\\
In this region absorption occurs due to only the excitation of particles to higher energy levels.
 As $k_3\longrightarrow 0$, then $k_{\parallel}^{2}=-\omega^2$.
  The solution for the energy are: $\varepsilon_{p}^{(1)}=\varepsilon_{p}^{(2)}=J_{nn^\prime}/2\omega$. In order to
   have positive energies, which could be important in the degenerate limit, the sum is restricted to $n^\prime>n$. Also, the condition $k_{\parallel}^{2}>k_{\parallel}^{2\,{\prime\prime}}$, in this limit, implies $2eB>\omega^2$.
 Finally, considering the degenerate limit, the Ohm conductivity in I is given by
\[
\sigma^{3D}_{0}(\omega,B,\mu,0)\hspace{-0.1cm}=\hspace{-0.1cm}\frac{e^{3}B}{8\pi\omega}\hspace{-0.07cm}\sum_{n=0}^{n_{max}}\hspace{-0.1cm}
\frac{2eB(2n+1)-\omega^{2}}{((2eB\hspace{-0.07cm}-\hspace{-0.07cm}\omega^2)^2\hspace{-0.07cm}-\hspace{-0.07cm}4\varepsilon_{0,n}^2\omega^2)^{1/2}}
\]
\begin{equation}
(\theta(\mu\hspace{-0.07cm}-\hspace{-0.07cm}\varepsilon_p^{(1)})\hspace{-0.07cm}-\hspace{-0.07cm}\theta(\mu\hspace{-0.07cm}-\hspace{-0.07cm}\varepsilon_p^{(1)}-\hspace{-0.07cm}\omega)).
 \label{eq:ohm3dexcitationexpandido}
\end{equation}

The sum over the integer $n$ goes to $n_{max}$ determined by the
restriction imposed by Eq. (\ref{detalles2}).  The combination of
the degenerate functions tells us that the Ohm conductivity does not
vanish if

$$\frac{2eB-\omega^2}{2\omega}<\mu<\frac{2eB+\omega^2}{2\omega}$$

\textbf{ b)} \textbf{Region II:} $k_{\parallel}^{2\,{\prime}}>k_{\parallel}^{2}$ pair creation.\\

In this region absorption may be due to excitation and also to pair creation. The
corresponding solution for the energies are:  $\varepsilon_{p}^{(3)}=\varepsilon_{p}^{(4)}=-J_{nn^\prime}/2\omega$.
In this case there is no restriction for the sum over n, and the
condition $k_{\parallel}^{2\,{\prime}}>k_{\parallel}^{2}$, implies $2eB<\omega^2$. Then, the expression
for the Ohm conductivity in the degenerate limit in II is
\begin{equation}
  \sigma^{3D}_{0}(\omega,B,\mu,0)\hspace{-0.1cm}=\hspace{-0.1cm}\frac{e^{3}B}{8\pi\omega}\hspace{-0.07cm}\sum_{n=0}^{n_{max}}
\hspace{-0.09cm}\frac{\omega^{2}\hspace{-0.07cm}-\hspace{-0.07cm}2eB(2n+1)}{((2eB\hspace{-0.07cm}-\hspace{-0.07cm}\omega^2)^2\hspace{-0.07cm}-\hspace{-0.07cm}4\varepsilon_{0,n}^2\omega^2)^{1/2}}\nonumber
\end{equation}
 \begin{equation}
 +\hspace{-0.1cm}\sum_{n=1}\hspace{-0.09cm}
\frac{\hspace{-0.1cm}\omega^{2}-2eB(2n-1)}{\hspace{-0.1cm}((2eB+\omega^2)^2-4\varepsilon_{0,n}^2\omega^2)^{1/2}}
(\theta(\varepsilon_p^{(3)}\hspace{-0.07cm}-\hspace{-0.06cm}\mu)+\theta(\omega\hspace{-0.06cm}-\hspace{-0.06cm}\varepsilon_p^{(3)}\hspace{-0.06cm}-\hspace{-0.07cm}\mu)), \label{eq:ohm3dpaircreationexpandido}\end{equation}

\noindent and from the degenerate distribution, we obtained that the
Ohm conductivity does not vanish if

\begin{equation}
 \frac{\omega^2+2eB}{2\omega}>\mu \quad or\quad
 \frac{\omega^2-2eB}{2\omega}>\mu.
 \end{equation}

Let us remark that in the static limit, $\omega=0$ the Ohm
conductivity is zero as was checked in Ref. \cite{aurora90}.

\section{Quantum Faraday Effect for a relativistic fermion gas}
\label{sec4}

Photons propagating in a relativistic fermion-antifermion
($e^{\pm}$) medium %at temperature
 at zero temperature and nonzero
particle density (chemical potential $\mu$) is of special interest
for astrophysics. The one-loop diagram describing the process
accounting for the photon self-energy interaction contains, in
addition to the virtual creation and annihilation of the pair, the
process of absorption and subsequent emission of one photon by the
fermions and/or antifermions.

The propagation of an electromagnetic wave in the medium can be
described by  the Maxwell equations
\begin{equation}
\partial_{\nu}F_{\nu\mu}+\Pi_{\mu\nu}a_{\nu}=0,\label{Maxwell3D}
\end{equation}

\noindent which could be written in momentum space as

\begin{equation}\label{dispersion}
\left[(k_{\perp}^2+k_{\parallel}^2)g_{\mu\nu}-k_{\mu}k_{\nu}+\Pi_{\mu\nu})\right]a_{\nu}=0,
\end{equation}

\noindent
 We will consider in what follows that  the photon propagates parallel to the magnetic field.
As in Sec. \ref{sec2},  $\Pi^{\nu\rho}$  is the self-energy of the
photon propagating in a magnetized dense medium, so it depends on
$T$, $\mu$, and magnetic field, apart from $k_3$ and $\omega$.

 To solve the dispersion relation (\ref{dispersion}) we need to diagonalize $\Pi_{\nu\rho}$. The
   general covariant structure of  the photon self-energy leads to the following
   expression:

    \begin{equation}
         \Pi_{\mu\nu}=\overset{3}{\underset{n=0}{\sum}}\kappa_{i}\frac{b_{{\mu}^{(i)}}b_{\nu}^{*(i)}}{b_{\mu}^{(i)}
            b_{\mu}^{*(i)}}, \quad\,\nu,\mu=1,2,3,4.
    \end{equation}

\noindent  where  $\kappa_{i}$ and $b_{\mu}^{(i)}$ are the
eigenvalues and the eigenvectors of $\Pi_{\mu\nu}$, respectively,
which satisfies the secular equation
        \begin{equation}
            \Pi_{\mu\nu}b_{\nu}^{(i)}=\kappa_{i}b_{\mu}^{(i)}.
        \end{equation}
In the particular case of propagation along the magnetic field
 there are three nonvanishing eigenvalues. The first
 two are transverse modes $b^{\prime (1,2)}_{\mu}$ \cite{hugo1982}-\cite{1} (see appendix for details).
\begin{equation}
b^{\prime (1,2)}_{\mu}=(b_{\mu}^{(1)}\pm
ib_{\mu}^{(2)}),\label{AMPO}
\end{equation}
\noindent with  $\quad\vec{b}_{\perp}^{(1)}=-\frac{\vec{k}_{\perp}}{k_{\perp}}\frac{k_{\parallel}}{k},
\quad b_{3,0}^{(1)}=0,$
\begin{equation}
b_{1}^{(2)}=\frac{k_{2}}{k_{\perp}} \quad
b_{2}^{(2)}=-\frac{k_{1}}{k_{\perp}},\quad
b_{3,0}^{(2)}=0,\end{equation} which describe  a circularly
polarized wave  in the plane perpendicular to \textbf{B} with
different eigenvalues,

\begin{equation}
\kappa_{1,2}=t\pm \sqrt{-r^2},  \label{autovalores}
\end{equation}
according to the definition given in  Eq. (\ref{integralIr}),
\begin{equation}
\kappa_{1,2}=t\pm I_r,  \label{autovalores}
\end{equation}
 that is, opposite directions, which
is the key of the Faraday effect. Also, there is a third mode
corresponding to a longitudinal wave which propagates along the
magnetic field $b_{\mu}^{3}$, and $\kappa_3=s$  is the corresponding
eigenvalue.

Let us consider the propagation of an electromagnetic wave, which at
$x_3=0$ is linearly polarized along the $x_1$ axes. Note that,
because the system has rotational symmetry with regard to \textbf{B}
( $k_{\perp}=0$), we can choose the direction of the eigenvectors
$b_{\mu}^{1,2}$ arbitrarily orthogonal to \textbf{B}. We can then
set $\frac{\vec{k}_{\perp}}{k_{\perp}}=\mathbf{e}_1$, and decompose
the wave into two circularly polarized waves
\begin{equation}
a_{\mu}=\displaystyle[\frac{1}{2}Ae^{i({k_{+}x_3-\omega
t})}b_{\mu}^{\prime (1)}+\frac{1}{2}Ae^{i({k_{-}x_3-\omega
t})}b_{\mu}^{\prime (2)}],
\end{equation}

\noindent where  $k_{\pm}$ are the solutions of the dispersion relations for the eigenmodes
 \begin{equation}
k_{\pm}=\displaystyle
\sqrt{\omega^{2}+\kappa_{1,2}}=\sqrt{\omega^{2}+t\pm I_r},\label{dispkappa}
\end{equation}

\noindent In order to solve the dispersion relation, the complex functions $r$ and $t$ in (\ref{dispkappa}) are considered in an approximation independent on $k_3$ (\ref{AMPO}).

The electric field associated with the wave is given by

 \begin{equation}\label{3D Faraday angle}
\mathbf{E}=\displaystyle[\frac{i\omega}{\sqrt{2}}Ae^{i({k_{+}x_3-\omega
t})}\mathbf{e}^{+}+\frac{i\omega}{\sqrt{2}}Ae^{i({k_{-}x_3-\omega
t})}\mathbf{e}^{-}],
\end{equation}
where $\mathbf{e}^{\pm}=(\mathbf{e}_1 \mp i \mathbf{e}_2)/\sqrt{2}$
are the polarization vectors of the left and right circularly
polarized waves, respectively. So, the superposition of both modes
leads to an elliptically polarized wave, whose principal axis
rotate.

The amount of the FR angle, after traveling a distance $L$ in the
medium, can be obtained from (see also Refs.
\cite{Avjit}-\cite{olivo})
  \begin{equation}
   \theta_{F}^{3D}=\frac{1}{2}\omega( n_{-}-n_{+})L,
\label{eq:faradayangleconindicesderefraccion}\end{equation}
  where $n_{\pm}$ are the refraction indices of the left- and right-circularly polarized waves, respectively, and can be defined as \cite{1}
  \begin{equation}
  n_{\pm}(\omega,k_3)=(1+\frac{\kappa_{1,2}(\omega,k_3)}{\omega^2})^{1/2}.
  \end{equation}
 Using $k=n\omega$, then the Faraday angle can be obtained directly from
Eqs. (\ref{3D Faraday angle}) and
(\ref{eq:faradayangleconindicesderefraccion}):
  \begin{equation}
\theta_{F}^{3D}=\frac{1}{2}(Re\left[k_{-}\right]-Re\left[k_{+}\right])L,
  \label{generalfaradyangle}\end{equation}
  where
  \begin{equation}
  \hspace{-0.15cm}
Re\hspace{-0.1cm}\left[k_{\pm}\right]\hspace{-0.12cm}=\hspace{-0.12cm}
\frac{\hspace{-0.05cm}1\hspace{-0.06cm}}{\hspace{-0.06cm}\sqrt{2}\hspace{-0.06cm}}\hspace{-0.12cm}\left[\hspace{-0.15cm}\sqrt{\hspace{-0.1cm}(\omega^{2}\hspace{-0.06cm}+\hspace{-0.06cm}Re[\kappa_{1,2}])^2\hspace{-0.06cm}+\hspace{-0.06cm}Im[\kappa_{1,2}]^2}\hspace{-0.06cm}+\hspace{-0.06cm}
 (\omega^{2}\hspace{-0.06cm}+\hspace{-0.06cm}Re[\hspace{-0.05cm}\kappa_{1,2}\hspace{-0.05cm}])\hspace{-0.12cm}\right]\hspace{-0.05cm}^{1/2}\hspace{-0.05cm}.
  \end{equation}
\noindent If $\text{Im}[\kappa_{1,2}]\ll\omega^{2}+Re[\kappa_{1,2}]$
we can roughly write
\begin{equation}
Re\left[k_{\pm}\right]\approx
\sqrt{\omega^{2}+Re[\kappa_{1,2}]}\left[1+\frac{Im[\kappa_{1,2}]^2}{4(\omega^{2}+Re[\kappa_{1,2}])^2}
 \right]^{1/2}.
  \end{equation}
Furthermore, if also $Re[\kappa_{1,2}]\ll\omega^{2}$, in the
leading-order approximation
\begin{equation}
Re\left[k_{\pm}\right]\approx
\omega+\frac{Re[\kappa_{1,2}]}{2\omega},
  \end{equation}and, according to
the relation given in Ref. (\ref{generalfaradyangle}), we obtain for
the rotated Faraday angle per unit length \footnote{We have returned
to the units $\hbar$ and $c$ to obtain this result.}
\begin{equation}
\frac{\theta_F^{3D}}{
L}\sim\frac{\sigma_H^{3D}}{2c}.\label{relacionHall1}
\end{equation}
Equation (\ref{relacionHall1}) shows the relation between Faraday
angle and the Hall conductivity. Let us note that in general the
Faraday angle depends on the terms of the admittivity tensor but the
leading term comes from the Hall conductivity. This result obtained
for 3D+1 systems  shows that it is a consequence of general
properties of QED in an external magnetic field at finite density.
In Sec. V we have obtained in 2D+1 limit. This result has been
obtained theoretically in 2D+1 systems \cite{Tan}, \cite{rusos},
\cite{japoneses1}.

\begin{figure}[h]
\begin{center}
 \includegraphics[width=8cm]{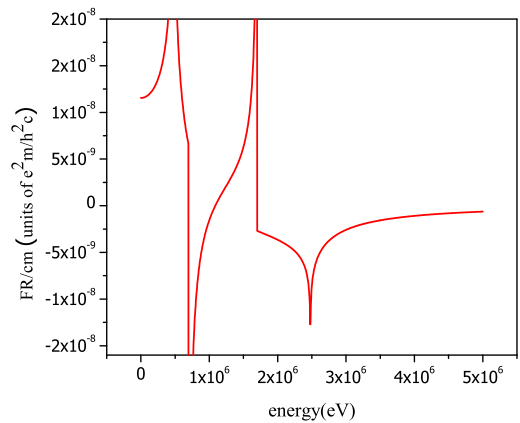}
 \caption{(Color online)Faraday angle per unit length as a function of  energy, for $\mu=1$ MeV and $B=10^{14}$ G,
 corresponding to $n_{\mu}$. The curve was plotted in a wide range of the photon energy $\hbar\omega=[10^2-5\times10^6]eV$ which includes the two branching points for the FR}\label{graficoangulo3d1}
\end{center}
\end{figure}

The Faraday angle in the degenerate limit($\sigma_H^{3D}$ is given
by Eq. (\ref{eq:3Dhallintegradaparan1})) has been depicted  in Fig.
\ref{graficoangulo3d1}, for  $n_{\mu}=0$ in a wide range of photon
energy. Because the Hall conductivity
(\ref{eq:3Dhallintegradaparan1}) has two branching points for
$n_{\mu}=0$, the curve has two peaks related to excitation
($\omega=-m+(m^2+2eB)^{1/2}$) and pair creation
($\omega=m+\sqrt{m^2+2eB}$). Then a resonant behavior for the
Faraday angle is obtained, associated with both absorption processes
\cite{hugo1982}. Let us note that the Faraday angle should be a
finite value. The divergences are avoided if we use the solution of
the dispersion equation near the singular points.

The relativistic quantized medium makes the angle depend nonlinearly
on the magnetic field,  contrary to the ¨classical¨ case of
interstellar medium where the relation with $B$ is linear and
depends on the electron density.

It is worthwhile to point out that Faraday effect is obtained as the
consequence of charge asymmetry of the system $\mu\neq0$. When the
system has charge symmetry  the  scalar $r$  vanishes  and the
Faraday effect is not manifested.

\section{Relativistic Hall and Ohm Conductivities in non-static limit ($\omega\neq 0$: 2D+1 system)}
\label{secrel} As is well known theoretically  properties of
graphene are essentially described by Dirac massless fermions
(electrons) in two dimensions. This system is ``relativistic" in the
sense that  the spectra of electrons and holes can be mimicked as
two-dimensional relativistic chiral fermions where electrons and
holes move at velocities $v_F\approx 10^6 m/s$ one hundredth the
speed of light \cite{NovoselovNature2005}. In this section with the
aim of studying a graphene like system we are going to obtain the 2D
Hall and Ohm conductivities in the nonstatic limit from the 3D
conductivities obtained in Sec. \ref{sec2}. Two considerations can
be made: the first is to do a dimensional compactification
\cite{aurora},\cite{aurora90} and the second one is to take the
limit $m\rightarrow 0$ \cite{aurora}. To consider the first of our
assumptions, we assume that the fermion-antifermion gas is confined
to a box of length $L_{3}$ and the limit $L_{3}\rightarrow0$ is
taken. Then the integral
 over $p_{3}$ is replaced by a sum
 over the integers $s=0,1,2,...$.
Because $p_{3}=2\pi s/L_{3}$ and $L_{3}\rightarrow0$ only the terms
$s=0$ remain in the sum. Then the 2D+1 limit is obtained taking
$p_{3}=0$ and $k_{3}=0$ and removing from all the expressions the
integrals $1/(2\pi\hbar)\int dp_{3}$. With this dimensional
reduction and the consideration of massless fermions in mind  for
2D+1 Hall conductivity at
 $\omega\neq 0$, we have ($Y_{ij}^{2D}=L_3Y_{ij}$)
\begin{equation}
\sigma_{H}^{2D}(\omega,B,\mu,T)\hspace{-0.1cm}=\hspace{-0.1cm}\frac{Im[ r^{2D}(\omega,B,\mu,T)]}{\omega}L_3,
\end{equation}
\noindent and
\[
\sigma_{H}^{2D}=\frac{e^{3}B}{\pi}\underset{n=0}{\overset{\infty}{\sum}}(\frac{(-\omega^2+2eB(2n+1))}{(-\omega^2+2eB)^{2}-
4\omega^{2}\varepsilon_{0,n}^{2}}\]
\begin{equation}
-\underset{n=1}{\overset{\infty}{\sum}}\frac{-\omega^{2}+2eB(2n-1)}{(\omega^{2}+2eB)^{2}-
4\omega^{2}\varepsilon_{0,n}^{2}})(n_{e}-n_{p})\label{Hall2dIr},
\end{equation}
\noindent where $\varepsilon_{0,n}=\sqrt{2neB}$. As in the earlier
section we consider the zero-temperature limit which means
substituting $n_e(\varepsilon)=\theta (\mu-\varepsilon)$  and zero
contribution of antifermions, since the gas is completely
degenerate. The Hall Conductivity has been written as
\begin{equation}
\sigma_{H}^{2D}(\omega,B,\mu,0)= \frac{e^{3}B}{\pi}\left\{ \frac{-\omega^{2}+2eB(2n_{\mu}+1)}{(2eB-\omega^{2})^{2}-4\omega^{2}\varepsilon_{0,n_{\mu}}^{2}}\right\}.\label{Hall2d}
\end{equation}
Let us remark that at $\omega=0$ we recover the expression of
quantum Hall Conductivity
$\sigma_{H}^{2D}=2\frac{e^{2}}{h}(n_{\mu}+\frac{1}{2})$.\footnote{we
have recovered the units $\hbar$ and $c$ to write this result.} The
Ohm conductivity should be obtained doing the dimensional reduction
in  Eq. (\ref{integralIt}) considering  massless fermions. In the
degenerate limit, we obtain
\begin{widetext}
\begin{equation}
  \sigma^{2D}_{0}(\omega,B,\mu,0)\hspace{-0.1cm}=\hspace{-0.1cm}\frac{e^{3}B}{2\pi\omega}
\sum_{n=0}^{n_{\mu}}\hspace{-0.1cm}\varepsilon_{0,n}\hspace{-0.1cm}\left(\frac{2\omega^2+4eB}{((-\omega^2+2eB)^2-4\omega^2\varepsilon_{0,n}^2)}
+\frac{2\omega^2-4eB}{((\omega^2+2eB)^2-4\omega^2\varepsilon_{0,n}^2)}\right)\theta(\mu-\varepsilon_{0,n})
 \label{eq:ohm2dexcitationexpandido},\end{equation}
\end{widetext}
Let us note that in Eq. (\ref{eq:ohm2dexcitationexpandido}) the sum
over $n$ goes to  $n_\mu=I[\frac{\mu^2}{2eB}]$.

Although our method of dimensional reduction  described above is
valid for getting 2D+1 limits,  it would be interesting to make a
full  2D+1 analysis of the problem by discussing the set of
independent tensor structures involved  and their relation to the
obtained results. Reference \cite{Zeitlin} is an early attempt to
deal with the 2D+1 case related to the Chern-Simons addition to the
Lagrangian.

\begin{figure}[h]
\begin{center}
 \includegraphics[width=8cm]{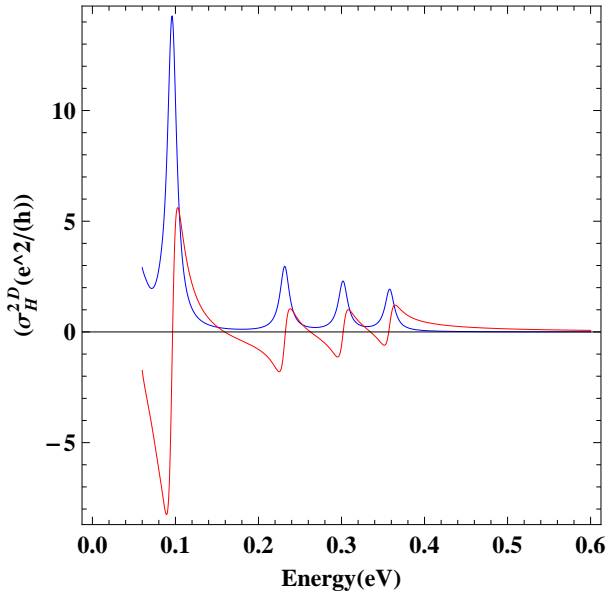} \caption{(Color online)O
 hm Conductivity (solid blue line) as a function of the photon energy for $B=7\times 10^4$ G, $\mu=200$ MeV and
  $\epsilon=6.8$ MeV. We use $v_f=10^8$ cm/s. We also have plotted the imaginary part of the conductivity (dashed red line).}\label{2dohmconductivity}
\end{center}
\end{figure}

To consider a graphene-like system in Eqs. (\ref{Hall2d}) and
(\ref{eq:ohm2dexcitationexpandido}),  additional considerations must
be taken into account. When the units $\hbar$ and $c$ are recovered,
$c\rightarrow v_f^2/c$ ($v_f$ is the Fermi velocity)\cite{aurora}.
The expressions for the Hall and Ohm conductivities [Eqs.
\ref{Hall2d})-(\ref{eq:ohm2dexcitationexpandido}] must be multiplied
by two, to account for the sublattice-valley degeneracy in graphene.
The frequency must be substituted by $\omega\rightarrow \omega
+i\epsilon$ where the imaginary part-$\epsilon$ is a
phenomenological parameter associated with system disorder
(\cite{rusos},\cite{castro}). In Fig. \ref{2dohmconductivity} the 2D
Ohm conductivity is plotted as a function of  energy for fixed
values of $B=7\times10^4$ G, chemical potential $\mu=200$ MeV and
$\epsilon=6.8$ MeV, which are typical values for a graphene-like
system \cite{castro} and \cite{rusos}. The figure also shows the
imaginary part of the conductivity. Our results obtained with the
\textbf{\emph{ansatz}} of a dimensional compactification are in
agreement with the theoretical studies of the Ohm conductivity in
graphene \cite{castro},\cite{gusynin2009}.

\section{2D+1 system: Faraday effect and  rotation Faraday angle}
\label{sec5} The purpose of this section is to study the Faraday
effect for a 2D+1 system (i.e., a graphene like system) starting
from the results obtained in Sec. \ref{sec4}. Let us suppose that
the graphene plate is located at $x_3=0$ and the incoming
electromagnetic wave is linearly polarized along the $x_1$
direction, and travels in the positive $x_3$ direction. Due to the
optical Faraday rotation of the polarization vector when the wave
crosses the graphene sheet, both the reflected and  transmitted
component acquire a component along the $x_2$ direction
(\cite{rusos},\cite{castro},\cite{wallace}).

We can formally follow the procedure of by \ref{sec4} by using the
solution of the dispersion relation for a photon in  a stratified
medium, given  by  a 3D+1 relativistic electron-positron plasma,
situated between $x_3=0$ and $x_3=L_3$,  in vacuum.

To describe the propagation of an electromagnetic wave in the whole
space, let us start from the modified Maxwell   Eq.
(\ref{Maxwell3D}) as
\begin{equation}
\partial_{\nu}F^{\nu\mu}+(\theta(x_3)-\theta(x_3-L_3))\Pi_{\mu\nu}a_{\nu}=0\label{maxwell2d}
\end{equation}
\noindent where the $\theta$-functions account for the inhomogeneity
in the Maxwell equation, which is only at $0<x_3<L_3$. The boundary
conditions at the  medium surfaces imply the continuity of the
electric field
\begin{equation}
E^{i}(x_3=0-,L_3-)=E^{i}(x_3=0+,L_3+)
\label{eq:matchesconditions1}\end{equation} and its
derivatives
\begin{equation} \partial_3 E^{i}(x_3=0-,L_3-)= \partial_3
E^{i}(x_3=0+,L_3+).
\label{eq:matchesconditions2}\end{equation}
  If we consider  an incident electromagnetic wave linearly polarized
along the $x_3$ direction,
\begin{equation}\label{wave inc}
\mathbf{E}^{I}=\displaystyle \frac{E}{\sqrt{2}}e^{i(k x_3-\omega
t)}\mathbf{e}^{+}+\frac{E}{\sqrt{2}}e^{i(k x_3-\omega
t)}\mathbf{e}^{-},
\end{equation}
the  transmitted wave ($x_3>L_3$) can be written as
\begin{equation}\label{wave trans}
\mathbf{E}^{T}=\displaystyle \frac{E^{+}}{\sqrt{2}}e^{i(k x_3-\omega
t)}\mathbf{e}^{+}+\frac{E^{-}}{\sqrt{2}}e^{i(k x_3 -\omega
t)}\mathbf{e}^{-},
\end{equation}
where $ E_{\pm}=(E_{1}\pm iE_{2})$ are complex amplitudes and
$\mathbf{e}^{\pm}=(\mathbf{e}_1 \mp i \mathbf{e}_2)/\sqrt{2}$
correspond to the left- and right-polarized waves, respectively
\cite{Jackson}. In order to express the amplitudes $ E_{\pm}$ in
terms of the medium parameters and the amplitude of the incident
wave $E$, we  can follow the multiple-reflections method described
in Ref. \cite{MaxBorn}. Let us define the complex total transmission
coefficients amplitudes $T_{\pm}=E_{\pm}^{T}/E_{\pm}^{I},$
\begin{equation}\label{transCoef}
T_{\pm}=\frac{\tau_{\pm} e^{ik_{\pm}L_3}}
{1+\varrho_{\pm}e^{2ik_{\pm}L_3}},
\end{equation}
where the factors $\tau_{\pm}= \displaystyle
\frac{4kk^{\pm}}{(k^{\pm}+k)^{2}},$ and $\varrho_{\pm}=
\frac{(k^{\pm}-k)^{2}}{(k^{\pm}+k)^{2}}$ come from the boundary
conditions (\ref{eq:matchesconditions1}) and
(\ref{eq:matchesconditions2}) and the exponentials $e^{ik_{\pm}L_3}$
are related to the FR due to the propagation in the medium  (as was
shown  in detail in Sec. III). Because $T_{\pm}$ are complex
numbers, they can be written as $T_{\pm}=|T_{\pm}|e^{i\theta_{\pm}
}$, and using the definition given above for the Faraday angle [Eq.
\ref{generalfaradyangle}]
 \begin{equation}
 \theta_{F}=\frac{1}{2}(\theta_{-}-\theta_{+}).
 \label{eq:faradaygraphene}\end{equation}
 In the limit case $k_{\pm}L_3\ll 1$, we can expand the exponentials
 in (\ref{transCoef}) up to the linear term in $L_3$, and obtain the approximate expressions
\begin{equation}
T_{\pm}\approx\displaystyle\frac{1}{1-i L_3
\frac{k_{\pm}^2+k^2}{2k}}=\frac{2}{1+L_3\frac{\kappa_{1,2}}{i\omega}-iL_3\omega}.
\end{equation}
 Finally, when $L_3\rightarrow 0$,
 $L_3\frac{\kappa_{1,2}}{i\omega}\rightarrow Y\pm=Y_{11}^{2D}\pm
 i Y_{12}^{2D}$, where $Y_{ij}^{2D}$ are the components of the 2D
 complex conductivity tensor, obtained from the 3D
 ones by the dimensional reduction prescription described  in the previous
 section. The Faraday angle in the 2D+1 limit is then given by
\begin{equation}
 \theta_{F}^{2D}=\frac{1}{2}(\theta_{-}^{2D}-\theta_{+}^{2D}),
 \label{eq:faradaygraphene}\end{equation}
where $\theta_{\pm}^{2D}=\arg{[T_{\pm}^{2D}]} $ and
\begin{equation}
    T_{\pm}^{2D}=\frac{2}{2+ Y_{\pm}}. \label{eq:solucionest}\end{equation}
It is easy to see from the  last two equations that, in the leading
order approximation,
\begin{equation}
 \theta_{F}^{2D}=\frac{1}{2}\sigma^{2D}_{H}.
 \end{equation}
This relation between the Faraday angle and the Hall Conductivity
has been already obtained in graphene \cite{japoneses},
\cite{rusos},\cite{castro} and here we have obtained it naturally
from the 3D result after a dimensional compactification.

It can be easily checked that our approach is equivalent to the one
followed in Ref. \cite{rusos} by taking the limit $L_3\rightarrow 0$
in Eq. (\ref{maxwell2d}).

\begin{figure}[h]
\begin{center}
 \includegraphics[width=8cm]{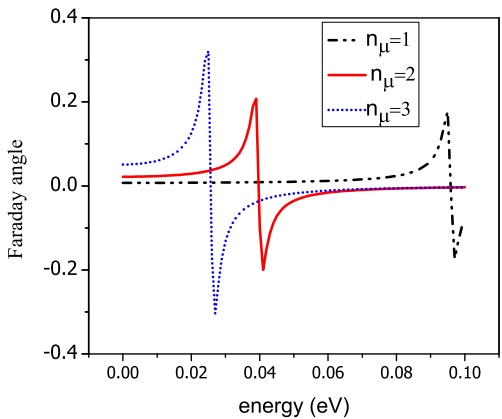}\caption{(Color online)
 Faraday angle as a function of the energy for $B=7\times 10^4$ G, $\epsilon=1$ MeV and three different
  values of the chemical potential: $\mu=(30, 110, 180)$ MeV}\label{graficoFaraday2dfrequency}
\end{center}
\end{figure}
\begin{figure}[h]
\begin{center}
 \includegraphics[width=8cm]{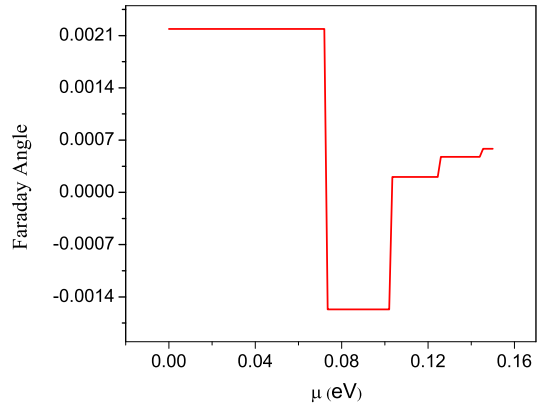} \caption{(Color online)
 Faraday angle as a function of the chemical potential for $B=4\times 10^4$ G, $\hbar\omega=150$ MeV and
  $\epsilon=0.5$ MeV. We use $v_f=10^8$ cm/s.}\label{graficoFaraday2dmiu}
\end{center}
\end{figure}
The Faraday angle is plotted in Figs.
(\ref{graficoFaraday2dfrequency}) and (\ref{graficoFaraday2dmiu}) in
the degenerate limit [in which Hall conductivity is given by  Eq.
(\ref{Hall2d})]. Figure (\ref{graficoFaraday2dfrequency}) shows the
Faraday rotation angle  versus $\omega$ for a fixed value of the
magnetic field $7\times 10^{4}$ G and  maximum Landau numbers
$n_{\mu}=0,1,3$, which corresponds to  chemical potentials-$\mu=(30,
110, 180)$ MeV, respectively. Each  curve shows  peaks associated
with  the poles of the Hall conductivity (\ref{Hall2d}) showing a
resonant behavior for the Faraday angle when the frequency reaches
the values corresponding to the poles and absorption processes
occurs. The curves were done assuming  $\epsilon=1$ MeV which is a
typical value of this quantity in graphene-like systems. The maximum
rotation angle for the parameters chosen is in agreement with the
angle predicted by Ref. \cite{Gusyin}. In Fig.
\ref{graficoFaraday2dmiu} the Faraday angle is plotted as a function
of chemical potential fixing $B=4\times 10^4$ G, $\hbar\omega=150$
MeV and $\epsilon=0.5$ MeV. The curve  shows a quantized behavior in
the same way as the Hall conductivity.

\section{Conclusions}
 \label{sec6}

The study of propagation of an electromagnetic wave parallel to a
magnetic field has been done starting from quantum field theory
formalism at finite temperature and density. The quantum Faraday
effect has been studied in 3D+1 and 2D+1 systems. We have obtained
the relation between the FR angle and Hall Conductivity (the Faraday
angle is given by the complex conductivity, but the leading term
comes from the Hall conductivity). Our finding shows that it is a
consequence of  general properties of propagation of an
electromagnetic wave parallel to a constant magnetic field, in a
quantum and relativistic dense system. We have found that Faraday
effect is consequence of the $C$ noninvariance of the system.

Due to the relation between Faraday effect and Hall conductivity we
 started our calculations studying the conductivity tensor in
the nonstatic limit. Hall and Ohm conductivities have been
calculated in the limit of zero temperature relevant for
applications to astrophysical and graphene-like systems. The
calculations can be extended to the general case of finite
temperature.

Let us remark that in the present paper we have focused on the real
Ohm and Hall  conductivities given by the imaginary part of $t$ and
the principal value of the integral $I_r$, respectively. A more full
discussion of the complex conductivity  including for instance the
imaginary part of integral $I_r$ (related to absorptive processes)
and the real part of $t$ will be discussed in a separate work.

The 2D+1 quantum relativistic system has also been studied by
introducing the ansatz of dimensional compactification of the $x_3$
dimension, allowing us to obtain the Hall and Ohm conductivities in
the limit of zero temperature.

The Faraday angle shows a quantized behavior in both 3D+1 and 2D+1
relativistic system. The dependence  of the Faraday angle with
frequency in graphene-like system is in agrement with theoretical
studies \cite{rusos}-\cite{castro}. Giant Faraday angles are found
for photon absorption frequencies. The outcome related to the
Faraday rotation angle in the astrophysical context, in particular
in the study of propagation of light in the magnetosphere of neutron
stars, deserves to be carefully analyzed. A separate study of this
topic will be addressed in a future work.

\section*{Acknowledgments}
The authors thanks to A. Cabo and A. Gonzalez Garcia for fruitful discussions. This work has been  supported  under the grant CB0407 and the ICTP Office of External Activities through NET-35.

\appendix
\section{Properties of the photon self-energy tensor}
\label{sec7}

  As a starting point we summarize some of the main features related
  with the photon self-energy of an electron-positron plasma in
  the presence of a constant magnetic field in the case of nonzero
  temperature and nonvanishing chemical potential.
  Under these conditions the polarization tensor may be expanded in
  terms of six independent transverse tensors \cite{hugo1982}. As is shown in Ref. \cite{1},
   symmetric properties in quantum statistics, corresponding to generalization of the Onsager relations, reduce
    the number of the basic tensors from 9 to 6:

 \begin{equation}\label{diagonalize form of the polarization tensor}
   \Pi_{\mu\nu}=\overset{6}{\underset{n=1}{\sum}}\pi^{(i)}\Psi_{\mu\nu}^{(i)},\quad\nu,\mu=1,2,3,4
 \end{equation}
The basic tensors are
\begin{eqnarray}
\Psi_{\mu\nu}^{(1)}\hspace{-0.1cm}&=&\hspace{-0.1cm}k^{2}g_{\mu\nu}-k_{\mu}k_{\nu},\quad\Psi_{\mu\nu}^{(2)}=F_{\mu\lambda}k^{\lambda}F^{\nu\rho}k^{\rho},\\
\Psi_{\mu\nu}^{(3)}\hspace{-0.1cm}&=&\hspace{-0.1cm}-k^{2}(g_{\mu\nu}\hspace{-0.06cm}-\hspace{-0.06cm}\frac{k_{\mu}k_{\lambda}}{k^{2}})F_{\rho}^{\lambda}F^{\rho\eta}(g_{\eta\nu}-\frac{k_{\eta}k_{\nu}}{k^{2}}),
\end{eqnarray}
\begin{widetext}
\begin{equation}
\Psi_{\mu\nu}^{(4)}=\frac{-[(F^{2}k)_{\mu}k^{2}-k_{\mu}(kF^{2}k)](F^{*}k)_{\nu}+(F^{*}k)_{\mu}[(F^{2}k)_{\nu}k^{2}-k_{\nu}(kF^{2}k)]}{(kF^{*2}k)[-k^{2}((kF^{2}k))]^{1/2}},
\end{equation}
\end{widetext}

\noindent where $g_{\mu,\nu}=(-1,1,1,1)$ is the metric tensor and
$F^{*}_{\mu\nu}$ is the dual of the electromagnetic tensor
$F_{\mu,\nu}$.

These tensors are symmetric in the indexes $\mu$, $\nu$ while the
following ones  are antisymmetric
\begin{eqnarray}
\Psi_{\mu\nu}^{(5)}\hspace{-0.07cm}&=&\hspace{-0.07cm}(u\cdot
  k)(k_{\mu}F_{\mu\lambda}k^{\lambda}-k_{\nu}F_{\mu\lambda}k^{\lambda}+k^{2}F_{\mu\nu}),\\
  \Psi_{\mu\nu}^{(6)}\hspace{-0.07cm}&=&\hspace{-0.07cm}u_{\lambda}F_{\mu\lambda}k^{\lambda}-u_{\nu}F_{\mu\lambda}k^{\lambda}-(u\cdot
  k)F_{\mu\nu}.
  \end{eqnarray}
 We introduce a set of orthonormal vectors which are the
  eigenvectors of $\Pi_{\mu\nu}$ in the limit $\mu=0$ and
  $\beta^{-1}=0$
\begin{eqnarray}
b_{\mu}^{(1)}&=&\frac{(F^{2}k)_{\mu}k^{2}-k_{\mu}(kF^{2}k)}{(-k^{2}(kF^{2}k)(kF^{*2}k))^{1/2}},\\
b_{\mu}^{(2)}&=&\frac{(F^{*}k)_{\mu}}{(kF^{*2}k)^{1/2}},\\
b_{\mu}^{(3)}&=&\frac{(Fk)_{\mu}}{(-kF^{2}k){}^{1/2}},\\
b_{\mu}^{(4)}&=&\frac{k{}_{\mu}}{(k^{2})^{1/2}}.
\end{eqnarray}
In the reference system in which the electron positron plasma is at
rest, the vectors $b_{\mu}^{(i)}$ look like
\begin{eqnarray}
\vec{b}_{\perp}^{(1)}=-\frac{\vec{k}_{\perp}}{k_{\perp}}\sqrt{\frac{k^{2}_{\parallel}}{k^{2}}},\quad
b_{3,0}^{(1)}=k_{3,0}\sqrt{\frac{k^{2}_{\perp}}{k^{2}_{\parallel}k^{2}}},\\
\vec{b}_{\perp}^{(2)}=0,\quad b_{3}^{(2)}=\frac{k_{0}}{\sqrt{k^{2}_{\parallel}}},\quad
b_{0}^{(2)}=\frac{k_{3}}{\sqrt{k^{2}_{\parallel}}},\quad\\
b_{1}^{(3)}=\frac{k_{2}}{\sqrt{k^{2}_{\perp}}},\quad
b_{2}^{(3)}=-\frac{k_{1}}{\sqrt{k^{2}_{\perp}}},\quad
b_{3,0}^{(3)}=0,\\
b_{\mu}^{(4)}=\frac{k_{\mu}}{(k^{2})^{1/2}}.\hspace{3cm}
\end{eqnarray}
 Using these vectors we can derive the scalars
\begin{eqnarray}
p&=&b^{(1)\mu}\Pi^{\nu}_{\mu}b^{(1)}_{\nu},\\
s&=&b^{(2)\mu}\Pi^{\nu}_{\mu}b^{(2)}_{\nu},\\
t&=&b^{(3)\mu}\Pi^{\nu}_{\mu}b^{(3)}_{\nu},\\
r&=&b^{(3)\mu}\Pi^{\nu}_{\mu}b^{(1)}_{\nu},\label{escalares}
\end{eqnarray}
\noindent and the pseudoscalar
\begin{eqnarray}
q&=&b^{(2)\mu}\Pi^{\nu}_{\mu}b^{(1)}_{\nu},\\
v&=&b^{(2)\mu}\Pi^{\nu}_{\mu}b^{(3)}_{\nu}.
\end{eqnarray}

In terms of this quantities the scalars $\pi^{(i)}$ in Eq.
\ref{diagonalize form of the polarization tensor} may be written in
the rest frame $(u_{\nu}=(0,0,0,1))$ as

\begin{eqnarray}
\pi^{(1)}&=&s/k^{2},\\
\pi^{(2)}&=&(k_{\parallel}^2t-k^{2}p+k_{\perp}^2s)/B^{2}k_{\parallel}^2k_{\perp}^2, \\
\pi^{(3)}&=&(p-s)/B^{2}k_{\parallel}^2,\pi^{(4)}=q, \\
\pi^{(5)}&=&-\frac{r}{B\omega(k^{2}k_{\parallel}^{1/2})}-\frac{v}{Bk_{3}(k_{\parallel}^2k_{\perp}^2)^{1/2}},\\
\pi^{(6)}&=&\frac{v}{B k_{3}}\frac{k_{\parallel}}{k_{\perp}}.
\end{eqnarray}

The above expression were written taking into account that the
magnetic field is directed along the $x_{3}$ axis. Then,
$\Pi_{\mu\nu}$ can be expressed in terms of the base vectors
$b^{i}_{\nu}$:
\begin{eqnarray}
\Pi_{\mu}^{\nu}b_{\nu}^{(1)}=p b_{\nu}^{(1)}+q b_{\mu}^{(2)}+r b_{\nu}^{(3)},\\
\Pi_{\mu}^{\nu}b_{\nu}^{(2)}=-q b_{\nu}^{(1)}+s b_{\mu}^{(2)}+v b_{\nu}^{(3)},\\
 \Pi_{\mu}^{\nu}b_{\nu}^{(3)}=-r b_{\nu}^{(1)}+v b_{\mu}^{(2)}+t b_{\nu}^{(3)},\\
 \Pi_{\mu}^{\nu}b_{\nu}^{(4)}=0.
 \end{eqnarray}
Finally the polarization tensor can be expressed in the
$b_{\nu}^{(i)}$ base as
\begin{equation}
          \Pi_{\mu\nu}=\left[\begin{array}{ccc}
                p & q & r\\
                -q & s & v\\
            -r & v & t\end{array}\right].\label{polarization}
            \end{equation}
From these results the eigenvalues could be determined  by finding
the modes of the wave propagating in the medium. In the  case of
propagation along the magnetic field, $k_{\perp}=0$, we obtain
$q=v=0$ and $p=t$ and Eq.  (\ref{polarization})  becomes
 \begin{equation}
          \Pi_{\mu\nu}=\left[\begin{array}{ccc}
                t & 0 & r\\
                0 & s & 0\\
            -r & 0 & t\end{array}\right],\label{polarization2}
            \end{equation}
\noindent with the eigenmodes $b^{\prime\,(1,3)}=b^{(1)}\pm
ib^{(3)}$ and $b^{\prime\,(2)}=b^{(2)}$ and eigenvalues
$\kappa^{(1,3)}=t\pm \sqrt{-r^2}$ and $\kappa^{(2)}=s$. Equation
(\ref{polarization2}) is equivalent to
\begin{equation}
          \Pi_{\mu\nu}=\left[\begin{array}{ccc}
               t & r& 0 \\
               -r & t&0\\
                0 &0& s
             \end{array}\right],\label{simanti}
            \end{equation}
\noindent and  the eigenmodes
$b^{\prime\,(1,2)}_{\mu}=b^{(1)}_{\mu}\pm ib^{(2)}_{\mu}$ and
$b^{\prime\,(3)}_{\mu}=b^{(3)}_{\mu}$ and eigenvalues
$\kappa_{1,2}=t\pm\sqrt{-r^2} $ and
$\kappa_{3}=s.$\footnote{Equation \ref{simanti} corresponds to the
Eq. (4.38) of Ref. \cite{olivo}. The eigenvalues $\kappa_{1,2}=t\pm
I_r$ are their $\pi_T\pm \pi_P$. The scalars $t$ and $I_r$
corresponds to $\pi_T$ and $\pi_P$, respectively.}

On the other hand, it follows from Eq. (\ref{simanti}) that we can
calculate the scalars $r$ and $t$t, which are given as
\begin{equation}\label{rr}
r(k\hspace{-0.1cm}\mid
\hspace{-0.1cm}A;\mu,\hspace{-0.1cm}\beta^{-})\hspace{-0.1cm}=\hspace{-0.1cm}-\frac{e^3B}{2\pi^2\beta}\hspace{-0.1cm}\sum_{p_4}\hspace{-0.1cm}\sum_{n,n^{\prime}}
\hspace{-0.1cm}\int\limits_{-\infty}^{\infty}\hspace{-0.2cm}\frac{dp_3C_{12,\,21}
}{(p_4^{\prime\,2}+\varepsilon_p^2)((p_4^{\prime}+k_4)^2+\varepsilon_p^2)}\end{equation}
\noindent where $C_{12,\,21}=\pm i[p_4(p_4^{\prime}+k_4)
+p_3(p_3^{\prime}+k_3)+m^2]F^3_{n,n^{\prime}}$ and
$F^3_{n,n^{\prime}}(\frac{k_{\perp}^2}{eB})=\mid
T_{n-1,n^{\prime}-1}\mid^2+ \mid T_{n-1,n^{\prime}}\mid^2$
\noindent and

\begin{equation}\label{tt}
t(k\hspace{-0.1cm}\mid
\hspace{-0.1cm}A;\mu,\hspace{-0.1cm}\beta^{-})\hspace{-0.1cm}=\hspace{-0.1cm}-\frac{e^3B}{2\pi^2\beta}\hspace{-0.1cm}\sum_{p_4}\hspace{-0.1cm}\sum_{n,n^{\prime}}
\hspace{-0.1cm}\int\limits_{-\infty}^{\infty}\hspace{-0.2cm}\frac{dp_3C_{11,\,22}}{(p_4^{\prime\,2}+\varepsilon_p^2)((p_4^{\prime}+k_4)^2+
\varepsilon_p^2)},
\end{equation}
\noindent and $C_{11,\,22}=p_4^{\prime}[p_4(p_4^{\prime}+k_4)
+p_3(p_3^{\prime}+k_3)+m^2]F^2_{n,n^{\prime}}$ and
$F^2_{n,n^{\prime}}(\frac{z_2}{eB})=\mid T_{n-1,n^{\prime}-1}\mid^2-
\mid T_{n-1,n^{\prime}}\mid^2,$
\noindent with
\begin{widetext}
\begin{equation}\label{laguerre}
  T_{n,m}(p,y)=\int e^{ipy}\Psi_n(x)\Psi_m(x+y)dx=(\frac{m!}{n!})^{1/2}(-\frac{y-ip}{\sqrt{2}})^{n-m}e^{-ipy-\frac{p^2+y^2}{4}}L_m^{n-m}(\frac{p^2+y^2}{2}),
\end{equation}
\end{widetext}

\noindent where $L_m^{n-m}$ are the generalized Laguerre
polynomials.

The sum $\sum_{p_4}$ is done using the Matsubara formalism where we
have
\begin{equation}\label{matsubara}
   \int\limits_{-\infty}^{\infty}\frac{dp_4}{2\pi}\rightarrow\frac{1}{\beta}\sum_{p_4},\quad p_{4}=\frac{(2n+1)\pi}{\beta},\,\,\,\, n=0,\pm1,\pm2,\ldots
\end{equation}

\noindent and the sum is carried out using the prescription
\begin{equation}\label{residuos}
    \frac{1}{\beta}\sum_{n}F(\ldots\frac{2n\pi}{\beta})=-\frac{1}{\beta}\sum_{p}f^{\pm}(\theta_p)Res\{F(\ldots\theta_p)\},
\end{equation}
where  $f^{\pm}=\pm\dfrac{i\beta }{1-e^{\mp i\beta\theta}}$ and
$\theta_p$ are the poles of  $F$.

\subsection{Calculation of $Im[t]$, $\sigma^0$ in 3D+1:  $k_\parallel^{2}<0$}

In order to solve the integrals $I_t$ and $I_r$ which have
singularities due to the denominator $D$, which can be written as
\begin{widetext}
\begin{equation}
D^{-1}=\frac{1}{8\varepsilon_{n^{\prime}}\varepsilon_{n}\omega}\left
(\frac{1}{\varepsilon_{n^{\prime}}-\varepsilon_{n}-\omega+i\epsilon}-\frac{1}{\varepsilon_{n^{\prime}}-\varepsilon_{n}+\omega+i\epsilon}
-\frac{1}{\varepsilon_{n^{\prime}}+\varepsilon_{n}-\omega+i\epsilon}+\frac{1}{\varepsilon_{n^{\prime}}+\varepsilon_{n}+\omega+i\epsilon}\right)\label{denominador},
\end{equation}
\end{widetext}

where we have added an infinitesimal positive imaginary part to
$\omega$, in order to take advantage of the relation
\begin{equation}\label{delta}
   \frac{1}{s-\omega-i\epsilon}=P\frac{1}{s-\omega}+i\pi\delta(s-\omega),
\end{equation}
to extract the imaginary and real part of the integrals. The first
pair of singularities  are related to excitation of particles to
higher energies and the second two are connected to the pair
creation. For $\varepsilon_{n^{\prime}}>\varepsilon_{n}$ the first
and the third of these denominators may vanish only for $\omega>0$
and the second and fourth if $\omega<0$. If
$\varepsilon_{n^{\prime}}<\varepsilon_{n}$  the opposite condition
hold.

The imaginary part of the denominator  $D$ of the integrals
(\ref{integralIr}) and (\ref{integralIt}) can be written  as
\cite{hugo1982}
\begin{widetext}
\begin{equation}\label{denominador0}
  Im [D^{-1}]=\pm \frac{\pi}{8\varepsilon_n\varepsilon_{n^{\prime}}\omega}\left (\delta(\varepsilon_{n^{\prime}}-\varepsilon_n\mp \omega)+\delta(\varepsilon_{n^{\prime}}-\varepsilon_n\pm \omega)-\delta(\varepsilon_{n^{\prime}}+\varepsilon_n\mp \omega)\right),
\end{equation}
\end{widetext}

To calculate the integral over $p_3$ we can use the formula
\begin{equation}
\int\limits_{-\infty}^{\infty}dp_3f(p_3)\delta(g(p_3))=\sum_m\frac{f(p_3^m)}{\mid
g^{\prime}(p_3^m)\mid},\label{formula}
\end{equation}

\noindent where $p_3^m$ are the roots of $g(p_3)=0$.
\begin{equation}
\mid
\frac{d}{dp_3}(g(p_3))\mid=\frac{\Lambda}{2\varepsilon_n^m\varepsilon_{n^{\prime}}^m}.\label{resultado}
\end{equation}

\end{document}